# λ-DNA through a plasmonic nanopore – What can be detected by means of Surface Enhanced Raman Scattering?

Aliaksandr Hubarevich[a†], Jian-An Huang[a†], Giorgia Giovannini[a,b†], Andrea Schirato[a,d],Yingqi Zhao[a], Nicolò Maccaferri[c], Francesco De Angelis[a], Alessandro Alabastri[e] and Denis Garoli[*a]

Engineered electromagnetic fields in plasmonic nanopores enable enhanced optical detection and their use in single molecule sequencing. Here, a plasmonic nanopore prepared in a thick nanoporous film is used to investigate the interaction between the metal and a long-chain double strand DNA molecule. We discuss how the matrix of nanoporous metal can interact with the molecule thanks to: i) transient aspecific interactions between the porous surface and DNA and ii) optical forces exerted by the localized field in a metallic nanostructure. A duration of interaction up to tens of milliseconds enables to collect high signal-to-noise Raman vibrations allowing an easy label-free reading of information from the DNA molecule. Moreover, in order to further increase the event of detection rate, we tested a polymeric porous hydrogel placed beneath the solid-state membrane. This approach enables a slowdown of the molecule diffusion, thus increasing the number of detected interactions by a factor of about 20.

During recent years, the interest in solid-state nanopores for single-molecule analysis has rapidly grown[1,2]. Such nanopores exhibit potential advantages compared to their biological counterparts[3,4], such as the tuning of the physical and chemical properties, as well as their compatibility with mass-production. Moreover, they can represent a new class of nanopore sensors, able to overcome the main limitations of current electrical readout techniques: low throughput and limited bandwidth[5]. In fact, the simultaneous multiplexed electrical readout from many thousands of nanopores is much less than trivial. To address these limitations, the combination of electrical recording and optical detection[6] is now under investigation. In this framework, plasmonic nanopores are extremely interesting because engineered highly confined electromagnetic (EM) fields generated by metallic nanostructures enable enhanced optical spectroscopies, local control over temperature, thermophoresis of molecules and ions to/from the sensor, and trapping of entities[5,7]. At the moment, the use of solid-state nanopore in single molecule sequencing, both in electrical and optical read-out schemes, is hindered by the extremely rapid translocation of biomolecules (in particular DNA and proteins) through nanopores. Much effort has been placed into solving this problem, including attempts to slow transport using optical tweezers[8–10]. It's now demonstrated that the high local optical field produced by the plasmonic nanostructures can be used to apply optical forces directly to nanoscale objects. In 2015 Belkin et al.[11] proposed and theoretically demonstrated single molecule DNA sequencing in a plasmonic nanopore. In that pioneering work, strongly localized EM fields are used to slow down the translocation of a DNA strand enabling at the same time the collection of Raman fingerprint spectra of the molecule.

Since its first demonstration in 1977[12], the methodology of surface-enhanced Raman scattering (SERS) has been improved to reach single-molecule sensitivity[13]. Importantly, Raman spectra report vibrational modes of the analyte molecules, providing direct information about their chemical structure. In the context of DNA sequencing, SERS signatures can be used to directly identify the four nucleotides of DNA without any labelling[14]. Recently, SERS detection of DNA single nucleotide translocation in a plasmonic nanopore has been reported[15]. Moreover, it has been demonstrated how SERS enables single nucleobase discrimination in single strand (ss)DNA in a plasmonic nanopore. This can be achieved thanks to the electro-plasmonic trapping effect, which allows to stably trap functionalized nanoparticles in a plasmonic nanopore[16]. Despite these intriguing results, literature on nanopore experiments on double strand (ds)DNA is still limited to a few preliminary works[17]. Again, the rapid translocation of the molecule through a solid-state nanopore represents the main limitation in this context, significantly reducing the maximum integration time in optical spectroscopies[18]. For these reasons, different methods are currently explored to extend the time duration of the interaction between the molecule and the plasmonic fields.

Here we experimentally translocate a λ-DNA, that is, a 48 kbp dsDNA[19], through plasmonic nanopore whileSERS spectra are collected. Several structures have been proposed as efficient plasmonic nanopores[5], here we use a simple fabrication method to prepare sub-20 nm apertures in a nanoporous gold (NPG) matrix. NPG is a well know plasmonic material[20,21] and its application for the development of SERS-based detection methods having sensitivity up to the single molecule,has been extensively proven[22–24]. In a first experiment we recorded for ten minutes (at intervals of 100 ms - integration time) the Raman spectrum emitted from λ-DNA molecules free to diffuse through the nanopore. We observed that the NPG pore enables to detect the main DNA fingerprint features. This suggests that the molecule can significantly interact with the metallic matrix and the localized EM filed by aspecific interactions and with not negligible optical forces. Moreover, it is clear that by reducing the velocity of translocation it should be possible to obtain spectroscopic information from the dsDNA employing a very simple platform. In this view, a nanoporous hydrogel has been integrated with the NPG pore. The use of agarose as polymeric matrix to slow DNA translocations has been recently proved[25,26] and here we want to demonstrate its application to a plasmonic latform.

Figure 1 illustrates the layouts of the platforms without (1b) and with (1c) the gel. The planar morphologies of the sample used for these experiments are also shown (1a). The hole prepared in the solid-state membrane was initially 25 nm in diameter, but the thick layer of metal deposited on it shrinks it down to sub 20 nm (see SI for details on the nanopore fabrication).

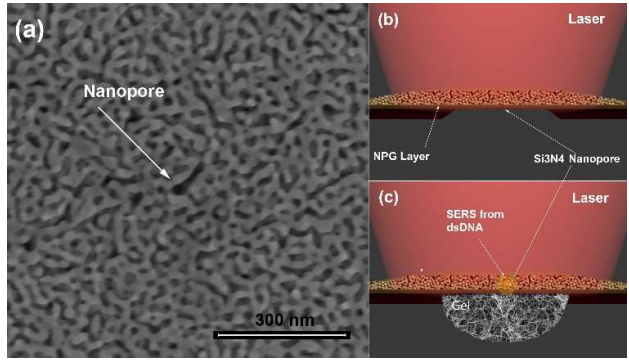

Figure 1: (a-c) SEM micrograph of the investigated sample and schematic layout of the structures.

We used an excitation wavelength of 785 nm to collect Raman spectrum of dsDNA from the nanopore prepared in the membrane. Two examples of the collected spectra are reported in Fig. 2. The peak identification can be performed considering the data reported in literature[27–30]. Importantly, within the 700 collected spectra, 45 of them show one or more peaks which can be easily ascribed to molecular vibrations of one of the four nucleobases. The discrete well-separated positions of the vibrational peaks collected during the experiment suggest that several DNA molecules passed through the nanopore sequentially. Considering the integration time of 100 ms, it seems that the molecules flowing through the nanopore interacted for tens of milliseconds with the NPG matrix. It's known that diffusing λ-DNA in solution is actually randomly coiled to maximize its entropy. A force of tens of pN is needed to stretch it[31]. Therefore, electric bias is usually needed to stretch the DNA molecule in order to make it pass through the nanopore uncoiled[9]. In this experiment, we did not apply an electric field to drive the molecule, but we let them freely diffuse through the pore (thanks to the different molar concentrations in two sides of the membrane).

The aspecific interaction between DNA and Au [32], combined with mechanical obstacle due to the nanoporous matrix and the localized optical field in the metal, can be considered as responsible for a reduced speed of translocation.

While the aspecific interaction is nontrivial to be assessed, the molecular diffusion and the optical force across the structure can be modeled and are here described. In particular, to provide a quantitative evaluation of their magnitude, Finite Element Method (FEM) models have been constructed via a commercial software (COMSOL Multiphysics 5.4). Both the models used an experimental NPG SEM micrograph (Fig. 3a) of the sample's cross section. The image was converted to a binary one with python opencv package as shown in Supporting Information (SI). As neither pressure nor electric potential were applied in the experiment, the diffusion process can be modeled by using the Flick's equation: $\partial c/\partial t + \nabla \cdot (-D\nabla c) = R$; where c is the concentration of the molecules, D is the diffusion constant of the molecules, and R is the source term. The diffusion constant was evaluated by Einstein-Smoluchowski relation (see SI for details). Figure 3b shows the concentration of the molecules going through the pore against time. The initial concentration at the bottom side of the chamber was set to 0.161mM, i.e. equal to the experimental λ-DNA concentration. As illustrated in Fig. 3b, even such a simple model demonstrates that NPG layer slows down the movement of molecules by around 30% during the first 100ms.

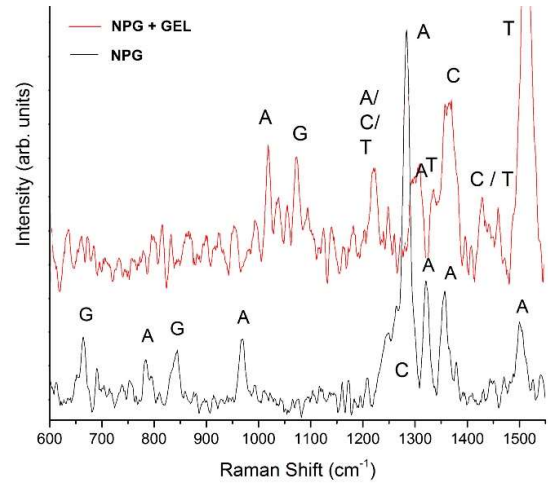

Figure 2: Examples of collected Raman spectrum (black curve represents a spectrum in the case of gel+NPG experiment; red curve is a spectrum in the case of NPG).

The effects of local optical forces on the molecule translocation has been simulated considering the interaction between an electromagnetic field and the NPG film. Also in this case the SEM image of the cross section of the experimental sample (Fig3a) is used directly to define the geometry in the model. The brightness levels of the SEM are exploited to encode and reproduce the porous morphology of the layer (Fig3c). Interestingly, this approach enables us to accurately account for the nanometric features of the structure, having a crucial role in the enhancement of the EM field and the localization of the optical force. Thus, once Maxwell's equations have been solved, the optical force acting on a dielectric sphere with refractive index $n_1$ can be evaluated according to[33] as: $\vec{F}_{opt} = 4\pi\varepsilon_0 n_2^2 r_A^3 (m^2 - 1/m^2 + 2)\vec{\nabla}\left(|\vec{E}|^2\right)$; where $|\vec{E}|$ is the magnitude of the EM field, $\varepsilon_0$ is the vacuum permittivity, $n_2$ the surrounding environment (here water) refractive index, $r_A$ the radius of the spherical molecule interacting with the field, $m = n_1/n_2$ its relative index. Fig3d shows then the spatial distribution of the norm $|\vec{F}_{opt}|$ when the system is excited at λ=785 nm, clearly exhibiting a relevant increase in the vicinity of the NPG-environment interface. The strength of the optical forces is thus enhanced, due to the field confinement induced by the plasmonic porous surface. To note, the maximum calculated force is about few fN, hence not enough to trap the molecule, but still effective to partially counterbalance thermal induced diffusion[34].

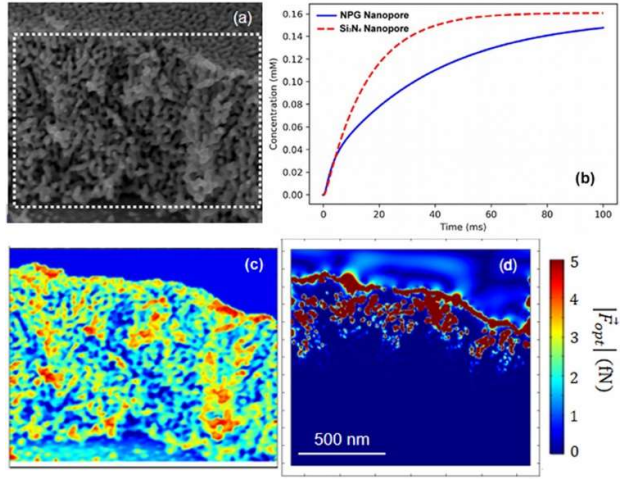

Figure 3: (a) SEM image of a cross section of the experimental NPG film (b) The average concentration of the molecules going through the pore against time. (c) The SEM image imported in the numerical FEM model (d) Norm of the optical forces theoretically computed across the cross section of he NPG at $\lambda = 785$ nm.

Data analysis on the experimental spectra have been performed using Python (https://www.python.org/). All spectra were selected in the 650-1500 cm$^{-1}$ range and smoothed using a Savitzky-Golay filter (see SI for more details). Data analysis allows to perform a statistical analysis on the obtained spectra. This has been performed by arbitrarily dividing the Raman shift energy range into 8 subsections: 640-680; 760-800; 1000-1040; 1060-1100; 1220-1260; 1280-1320; 1340-1380 and 1460-1500 cm$^{-1}$. An unambiguous assignment of the different Raman modes is not trivial. The exact position of the Raman shift depends on the orientation of the molecule and on the interaction between the molecule and the EM field. For these reasons, in literature, different values are assigned to the vibrational spectrum of DNA backbone, adenine (A), cytosine (C), guanine (G) and thymine (T)[14,30,27]. Table 1 reports the main peaks assignments used in our experiment.

| Raman Shift Range (cm$^{-1}$) | Main Peaks (cm$^{-1}$) | Assignments [ref] |
| --- | --- | --- |
| 640-680 | 656; 652; 689 | 5-ring and 6-ring deformation |
| 760-800 | 776; 792; 798 | ring breath and 6-ring deformation |
| 1000-1040 | 1026; 1034 | rock NH$_3$; wag CH$_3$ |
| 1080-1140 | 1140; 1150 | bend N9-H, C8-H; stretch N1-C2; C2-N3 |
| 1220-1260 | 1220; 1229; 1240 | bend C8-H, N10-H11; ring stretch C-N |
| 1280-1320 | 1290; 1307 | ring stretch C-N, C-C |
| 1340-1380 | 1338; 1370; 1385 | ring stretch C-N, C-C; rock NH$_3$ |
| 1460-1500 | 1461; 1490; 1482 | ring stretch C-N; bend C8-H; stretch C4-N8 |

As it can be seen from Fig. 2, the adenine (A) peaks are the easiest to be detected due to the large Raman cross section of the nucleotide[35] as well as its strongest adsorption on gold[32]. It can be detected in all the 45 significant spectra collected. The bands ascribable to A residue are at 792, 965, 1338, 1371 and 1490 cm$^{-1}$ [30]. We identify bands mainly due to the guanine (G) residues at 665/677 and 1050 cm$^{-1}$. The peak close to 850 cm$^{-1}$ can be related to 5-ring deformation of G or ring deformation of T[30]. Based on the analysis of the vibrational multiple peaks, bands can be detected between 1250 and 1500 cm$^{-1}$. They can be ascribable to the different nucleic acids; the intense feature at 1280 cm$^{-1}$ can be related to A or T residues, while the peak at 1520 cm$^{-1}$ is probably related to the in-plane ring stretch of T. In a dsDNA nanopore experiment we expect also to observe backbone related bands. In general, analysing the backbone modes enables to obtain information about the DNA chain conformation[36]. Here, we are not interested in this aspect since we probed a well-known dsDNA molecule from the bacteriophage $\lambda$[19]. In our experimental spectra we monitored the 1421 cm$^{-1}$ (DNA backbone vibrations) and the 1090 cm$^{-1}$ v(PO$_2$) bands. Only the second one can be detected, while the peak at 1421 cm$^{-1}$ seems to have a too low cross-section for our system.

Figure 4 reports the statistical analysis of the number of detected peaks within specific energy ranges. This shows that in our first experiment (Figs. 4a and b), at every event of detection, we can collect up to 11 different peaks. The largest number of peaks fall in the range between 1280 and 1380 cm$^{-1}$. This region contains many overlapping bands, mainly arising from the in-plane vibrations of base residues.

In a second experiment (Fig. 4c and 4d), a 1% agarose hydrogel has been prepared and integrated at the bottom of the nanoporous membrane (see SI for details). The hydrogel is known to reduce the velocity of translocation up to 1 order of magnitude[25,26,37]. Fig. 4c and 4d illustrate the results obtained from the SERS data collection by using the same parameters discussed above. Surprisingly, the number of events of detection of one or more peaks ascribable to DNA residues increased by a factor above 20.

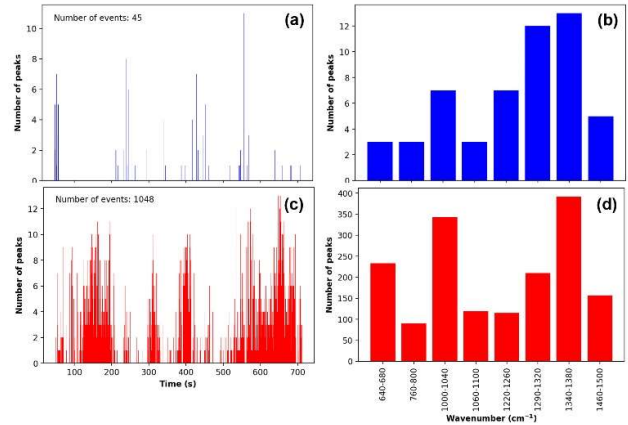

Figure 4: Statistical analysis of the detected peaks with and without the Agarose Gel. (a), and (c) are the event rate change with time, the information of which are extracted from a time trace. They mean that the event can happen more frequently with time. (b) and (d) are the histograms considering the spectral ranges of analysis.

The largest number of peaks was detected in the spectral range between 1000-1040 cm$^{-1}$ and 1340-1380 cm$^{-1}$. This is in accordance with the results obtained without the hydrogel. On the contrary, a higher number of events in the range between 640 up to 680 cm$^{-1}$ were observed.

In summary, we have shown that NPG can provide a longer and stronger foundation to further investigate the methodologies needed to optimize the control and to slow the translocation rate of DNA through the nanopore structures. Such an approach enables the measurement of the DNA molecule nucleobases. Indeed, when the latter translocate from the nanofluidic conduit channels through the nanoporous structure of the chip, the spectral visualization of the DNA base pair signatures using SERS spectroscopy becomes possible. As observed in the measured spectra, the translocating DNA within the nanopore barrier shows spectral bands that can be related and potentially assigned to the spectral signatures of the nucleobases of DNA. We also verified how the porous matrix can reduce the diffusion time and introduce not negligible optical forces. Finally, an hydrogel has been applied to the platform in order to significantly increase the number of detection events.

## Notes and references

1  Y. Feng, Y. Zhang, C. Ying, D. Wang and C. Du, *Genomics, Proteomics Bioinforma.*, 2015, **13**, 4–16.
2  Y. Goto, R. Akahori, I. Yanagi and K. Takeda, *J. Hum. Genet.*, , DOI:10.1038/s10038-019-0655-8.
3  C. Cao and Y. T. Long, *Acc. Chem. Res.*, 2018, **51**, 331–341.
4  Y. L. Ying, C. Cao and Y. T. Long, *Analyst*, 2014, **139**, 3826–3835.
5  D. Garoli, H. Yamazaki, N. Maccaferri and M. Wanunu, *Nano Lett.*, 2019, acs.nanolett.9b02759.
6  S. Huang, M. Romero-Ruiz, O. K. Castell, H. Bayley and M. I. Wallace, *Nat. Nanotechnol.*, 2015, **10**, 986–991.
7  J. D. Spitzberg, A. Zrehen, X. F. van Kooten and A. Meller, *Adv. Mater.*, 2019, **31**, 1–18.
8  U. F. Keyser, *J. R. Soc. Interface*, 2011, **8**, 1369–1378.
9  U. F. Keyser, B. N. Koeleman, S. van Dorp, D. Krapf, R. M. M. Smeets, S. G. Lemay, N. H. Dekker and C. Dekker, *Nat. Phys.*, 2006, **2**, 473–477.
10  L. Galla, A. J. Meyer, A. Spiering, A. Sischka, M. Mayer, A. R. Hall, P. Reimann and D. Anselmetti, *Nano Lett.*, 2014, **14**, 4176–4182.
11  M. Belkin, S. Chao, M. P. Jonsson, C. Dekker and A. Aksimentiev, *ACS Nano*, 2015, 10598–10611.
12  D. L. Jeanmaire and R. P. Van Duyne, *J. Electroanal. Chem. Interfacial Electrochem.*, 1977, **84**, 1–20.
13  E. C. Le Ru and P. G. Etchegoin, *Annu. Rev. Phys. Chem.*, 2012, **63**, 65–87.
14  S. Dick and S. E. J. Bell, *Faraday Discuss.*, 2017, **205**, 517–536.
15  C. Chen, Y. Li, S. Kerman, P. Neutens, K. Willems, S. Cornelissen, L. Lagae, T. Stakenborg and P. Van Dorpe, *Nat. Commun.*, 2018, **9**, 1733.
16  F. Huang, Jian-An; Mousavi, Mansoureh Z.; Zhao, Yingqi; Hubarevich, Aliaksandr; Omeis and F. Giovannini, 5 Giorgia; Schütte, Moritz; Garoli, Denis; De Angelis, *Nat. Commun.*, DOI:10.1038/s41467-019-13242-x.
17  E. A. Mendoza, A. Neumann, Y. Kuznetsova, S. R. J. Brueck and J. Edwards, *Opt. Laser Technol.*, 2019, **109**, 199–211.
18  T. Gilboa, C. Torfstein, M. Juhasz, A. Grunwald, Y. Ebenstein and E. Weinhold, , DOI:10.1021/acsnano.6b04748.
19  S. R. Casjens and R. W. Hendrix, *Virology*, 2015, **479**–**480**, 310–330.
20  D. Garoli, E. Calandrini, A. Bozzola, A. Toma, S. Cattarin, M. Ortolani and F. De Angelis, *ACS Photonics*, 2018, **5**, 3408–3414.
21  X. Zhang, Y. Zheng, X. Liu, W. Lu, J. Dai, D. Y. Lei and D. R. MacFarlane, *Adv. Mater.*, 2015, **27**, 1090–1096.
22  C. Fang, J. G. Shapter, N. H. Voelcker and A. V. Ellis, *RSC Adv.*, 2014, **4**, 19502–19506.
23  H. Liu, L. Zhang, X. Lang, Y. Yamaguchi, H. Iwasaki, Y. Inouye, Q. Xue and M. Chen, *Sci. Rep.*, 2011, **1**, 1–5.
24  L. Zhang, X. Lang, A. Hirata and M. Chen, *ACS Nano*, 2011, **5**, 4407–4413.
25  M. Waugh, A. Carlsen, D. Sean, G. W. Slater, K. Briggs, H. Kwok and V. Tabard-Cossa, *Electrophoresis*, 2015, **36**, 1759–1767.
26  N. Kaji, M. Ueda and Y. Baba, *Biophys. J.*, 2002, **82**, 335–344.
27  L. Guerrini, Ž. Krpetić, D. van Lierop, R. A. Alvarez-Puebla and D. Graham, *Angew. Chemie Int. Ed.*, 2015, **54**, 1144–1148.
28  V. S. Gorelik, A. S. Krylov and V. P. Sverbil, *Bull. Lebedev Phys. Inst.*, 2014, **41**, 310–315.
29  C. Otto, T. J. J. van den Tweel, F. F. M. de Mul and J. Greve, *J. Raman Spectrosc.*, 1986, **17**, 289–298.
30  F. Madzharova, Z. Heiner, M. Gühlke and J. Kneipp, *J. Phys. Chem. C*, 2016, **120**, 15415–15423.
31  T. R. Strick, M.-N. Dessinges, G. Charvin, N. H. Dekker, J.-F. Allemand, D. Bensimon and V. Croquette, *Reports Prog. Phys.*, 2002, **66**, 1–45.
32  K. M. Koo, A. A. I. Sina, L. G. Carrascosa, M. J. A. Shiddiky and M. Trau, *Anal. Methods*, 2015, **7**, 7042–7054.
33  Y. Harada and T. Asakura, *Opt. Commun.*, 1996, **124**, 529–541.
34  C. T. Yavuz, J. T. Mayo, W. W. Yu, A. Prakash, J. C. Falkner, S. Yean, L. Cong, H. J. Shipley, A. Kan, M. Tomson, D. Natelson and V. L. Colvin, *Science (80-. ).*, 2006, **314**, 964–967.
35  E. Papadopoulou and S. E. J. Bell, *Chem. - A Eur. J.*, 2012, **18**, 5394–5400.
36  D. Kurouski, T. Postiglione, T. Deckert-Gaudig, V. Deckert and I. K. Lednev, *Analyst*, 2013, **138**, 1665.
37  T. D. Howard and G. Holzwarth, *Biophys. J.*, 1992, **63**, 1487–1492.